\documentclass[12pt,preprint]{aastex}

\shorttitle{}
\shortauthors{Nesvorn\'y et al.}

\begin{document}
\baselineskip 19.pt

\title{Do Planetary Encounters Reset Surfaces of Near Earth Asteroids?}

\author{David Nesvorn\'y$^{1}$, William F. Bottke$^{1}$, David Vokrouhlick\'y$^2$, Clark R. Chapman$^{1}$
\and Scot Rafkin$^{1}$}
\affil{(1) Department of Space Studies, Southwest Research Institute,
1050 Walnut St., Suite 300, Boulder, Colorado 80302, USA}
\affil{(2) Institute of Astronomy, Charles University, V Hole\v{s}ovi\v{c}k\'ach 2, 
180 00 Prague 8, Czech Republic}

\begin{abstract}
Processes such as the solar wind sputtering and micrometeorite impacts can modify optical properties
of surfaces of airless bodies. This explains why spectra of the main belt asteroids, exposed to these
`space weathering' processes over eons, do not match the laboratory spectra of ordinary chondrite (OC) 
meteorites. In contrast, an important fraction of Near Earth Asteroids (NEAs), defined as Q-types in 
the asteroid taxonomy, display spectral attributes that are a good match to OCs. Here we study the possibility 
that the Q-type NEAs underwent recent encounters with the terrestrial planets and that the tidal gravity 
(or other effects) during these encounters exposed fresh OC material on the surface (thus giving it the 
Q-type spectral properties). We used numerical integrations to determine the statistics of encounters of
NEAs to planets. The results were used to calculate the fraction and orbital distribution of Q-type asteroids 
expected in the model as a function of the space weathering timescale, $t_{\rm sw}$ (see main text for 
definition), and maximum distance, $r^*$, at which planetary encounters can reset the surface. We found that 
$t_{\rm sw}\sim10^6$ yr (at 1 AU) and $r^* \sim 5$ $R_{\rm pl}$, where $R_{\rm pl}$ is the planetary radius, 
best fit the data. Values $t_{\rm sw}<10^5$~yr would require that $r^*>20$ $R_{\rm pl}$, which is probably 
implausible because these very distant encounters should be irrelevant. Also, the fraction of Q-type NEAs would 
be probably much larger than the one observed if $t_{\rm sw} > 10^7$ yr. We found that $t_{\rm sw} 
\propto q^2$, where $q$ is the perihelion distance, expected if the solar wind sputtering controls 
$t_{\rm sw}$, provides a better match to the orbital distribution of Q-type NEAs than models with fixed 
$t_{\rm sw}$. We also discuss how the Earth magnetosphere and radiation effects such as YORP can influence 
the spectral properties of NEAs. 
\end{abstract}


\section{Introduction}

Measurements of the spectral properties of Near Earth Asteroids (NEAs) provide important evidence 
concerning the relationship between asteroids and the most common class of meteorites known as the 
ordinary chondrites (OCs). The tendency toward seeing OC-like spectral attributes among NEAs has been 
noted in multi-filter color observations (Rabinowitz 1998, Whiteley 2001), and in visible and 
near-infrared specrophotometric surveys (Binzel et al. 1996, 2004, 2010). In contrast, no spectral 
analogs of OCs have been found to date among the $\sim$2000 surveyed main belt asteroids (MBAs), except 
for a case related to identified recent asteroid collisions (Moth\'e-Diniz and Nesvorn\'y 2008).

The lack of spectrophotometric analogs for OC meteorites in the main belt is a long-debated 
and fundamental problem. It is now generally accepted that processes similar to those acting on 
the Moon, such as solar wind sputtering and micrometeorite impacts (Gold 1955, Pieters et al. 2000, 
see Hapke 2001 and Chapman 2004 for reviews), can darken and redden the initially OC-like spectrum of a fresh 
asteroid surface, giving it the `weathered' appearance (see Chapman 1996 and Clark et al. 2001, 2002a,b 
for direct evidence for asteroid space weathering processes from the NEAR-Shoemaker and 
Galileo spacecrafts). In the following text we will refer to processes that alter optical properties 
of surfaces of airless bodies as the `space weathering' (SW) effects.   

Since the SW processes should affect the MBAs and NEAs in roughly the same way (see, e.g., 
Marchi et al. 2006 for a study of the SW dependence on heliocentric distance), it may seem puzzling
why a significant fraction of NEAs has an unweathered appearance (Binzel et al. 2004) while 
practically all spectroscopically surveyed MBAs are weathered. Several explanations have been proposed. 

To simplify the discussion of different models described below, we will use the following 
terminology taken from the standard asteroid taxonomy (Bus and Binzel 2002, DeMeo et al. 2009). 
We will define three categories of asteroid spectra: (1) Q-type spectra with deep absorption 
bands and shallow spectral slope similar to that of most OC meteorites in the RELAB 
database\footnote{http://www.planetary.brown.edu/relab/}; 
(2) S-type spectra with shallow absorption bands and relatively steep spectral slope
similar to that of weathered OCs; and (3) Sq-type spectra as the intermediate case 
between S and Q. See Bus and Binzel (2002) and  DeMeo et al. (2009) for a formal definition of these 
categories. We will assume that asteroids with the Q-type spectra have essentially unweathered surfaces 
with OC-like mineralogy; the Sq- and S-type asteroids will be assumed to have moderately and strongly 
weathered surfaces with OC-like mineralogy. 

About 20\% of chondritic NEAs surveyed at visible and/or near-infrared wavelengths have Q-type spectra while 
this fraction is essentially zero among MBAs. The standard interpretation of these results has been 
based on the presumption that Q-type asteroids are likely to be small. Indeed, the current spectrophotometric 
surveys of MBAs are largely incomplete in the size range of typical NEAs with diameters less then a few km.
Special emphasis has therefore been given to asteroid-size-dependent processes, such as immaturity of 
regoliths on small asteroids and/or the shorter collisional lifetime of smaller asteroids (e.g., Johnson 
and Fanale 1973, Binzel et al. 1996, 2004, Rabinowitz 1998, Whiteley 2001). 

One possibility is that the observed spectral variations may be related to particle-size effects 
(Johnson and Fanale 1973), where the decreasing gravity results in a different size distribution of 
surface particles on typically smaller NEAs than on larger MBAs. However, the photometric parameters 
indicative of particle-size effects show little evidence of an asteroid diameter dependence, thereby 
giving doubt to this explanation (e.g., Clark et al. 2001, Masiero et al. 2009).

Binzel et al. (2004) hypothesized that the SW size-dependency was because the survival lifetime against
catastrophic disruption decreases with decreasing size. Thus, on average, as we examine smaller and smaller 
objects, we should see younger and younger surfaces. Surfaces showing Q-type spectral properties should thus 
exist, on average, only among the smallest asteroids, which become easy spectroscopic targets only when they 
enter into NEA space. Large, OC-like asteroids in the main belt should have, on average, space-weathered 
spectral properties, explaining why they are taxonomically classified as S types.

\section{NJWI05 Model}

Nesvorn\'y et al. (2005; hereafter NJWI05) pointed out several problems with this ``standard model''. 
For example, the estimates of SW rates (Chapman et al. 2007, Moth\'e-Diniz and Nesvorn\'y 2008, Vernazza 
et al. 2009) suggest that SW probably operates on timescales $<$10 My (or perhaps even $\lesssim$1 My) to 
change an initially fresh Q-type surface into one that is partially weathered (corresponding to Sq). 
The SW process then probably continues to approach full maturity only after Gys of evolution (Jedicke 
et al. 2004, NJWI05, Willman et al. 2008). If so, the standard model would imply that the observed 
Q-type NEAs should have surface ages that are $<$10 My.

This implication of the standard scenario is at odds with the collisional and dynamical models of the 
NEAs' origin  because it would imply that $\sim$20\% of chondritic NEAs were produced by collisional 
breakups of large bodies within the past $<$10 My. In contrast, models predict much longer durations for 
processes like the Yarkovsky effect and weak resonances to insert km-sized MBAs into the planet-crossing space 
(e.g., Migliorini et al. 1998, Bottke et al. 2002, Morbidelli and Vokrouhlick\'y 2003), and collisional 
lifetimes $>$100 My (e.g., Bottke et al. 2005). Whiteley (2001) discussed additional objections to the 
standard model. 

More recently, as a test of the standard model, P. Vernazza and Moth\'e-Diniz et al. (2009) conducted a 
spectroscopic survey of $\approx$100 diameter $D\lesssim5$ km asteroids in the inner main belt. With only one 
possible (but uncertain) Q-type candidate detected (Moth\'e-Diniz et al. 2009), this survey indicates that
the Q-type asteroids are rare even among small MBAs.  This rules out the standard scenario that attempts to 
explain spectrophotometric differences between NEAs and MBAs as chiefly due to size-dependent effects. 

To resolve these problems, NJWI05 proposed a new model for the origin of Q-type NEAs by postulating that 
the optically-active layer on their surface has been recently reset by the {\it effects of tidal gravity during 
encounters of these bodies to the terrestrial planets.} For example, applied tidal stresses applied may cause 
elements of a fractured body to move with respect to each other, ballistically displace surface material, or even 
liberate the surface layers from the asteroid. Alternatively, if the tidal torque spins up an asteroid, the weathered 
regolith layers can be removed by carrying away the excess angular momentum.

To show the plausibility of this idea, NJWI05 
estimated that a typical NEA suffers on average about one encounter to within 2 Roche radii ($R_{\rm Roche}$) 
from the Earth every $\approx$5~My. This time interval between encounters is comparable with the average orbital 
lifetime of NEAs ($\approx$5~My according to Bottke et al. 2002) and is also comparable with the range of the 
SW timescale discussed above. Consequently, if tidal encounters at $2 R_{\rm Roche}$ can reset 
surfaces, Q-type NEAs should be numerous. For comparison, encounters up to five planetary radii (this 
limit depends on the NEA's shape and spin vector) can produce strong shape distortions of a rubble-pile NEA and 
material stripping up to 10\% of the NEA's pre-encounter mass (Richardson et al. 1998). 

Strong evidence supporting the NJWI05 model comes from the observed orbital distribution of Q-type NEAs. 
Several trends were pointed out (Nesvorn\'y et al. 2004, NJWI05, Marchi et al. 2006): (a) the proportion of 
Q-type NEAs increases with decreasing perihelion distance $q$; (b) the orbital distribution of known Q-type NEAs 
has a sharp edge at $q\sim1$ AU; and (c) concentrations of Q-type NEAs occur for values of $q$ that correspond 
to large collision probability with the terrestrial planets ($q=1.0$ and 0.72 AU for Earth and Venus; see, e.g., 
Morbidelli and Gladman 1998). This correlation of the orbital distribution of Q-type NEAs with the collision 
probability is expected in the NJWI05 model because orbits with large collision probability also have frequent 
close encounters with the terrestrial planets (e.g., Bottke \& Melosh 1996). Thus, the surfaces of these objects are 
expected to be `fresher' on average showing fewer signs of the SW effects. 

Recently, Binzel et al. (2010; hereafter B10) presented results supporting the NJWI05 model. They used a taxonomic 
classification based on the near-infrared (near-IR) spectroscopy which should more closely characterize 
surface mineralogy than previous taxonomies based on the visible wavelengths. We have verified, however, that there 
exists a very good correlation between the visible and near-IR classification of Q types. For example, from 6 NEAs 
that were classified as Qs from the visible spectra before IR data became available, 5 were classified as Qs 
based on the near-IR spectra and only 1 must have been re-classified as Sq. This shows that the use of the 
near-IR data does not really change the problem.

B10 numerically integrated the orbits of 95 selected asteroids (Q-, Sq-, and S-type NEAs and more distant 
Mars-crossers) for 0.5 My, recorded the history of their close encounters to the Earth and statistically analyzed 
the encounters in an attempt to correlate the statistics with spectroscopic type. Two main results were 
obtained in B10: 

(1) 20 Q- and 55 S/Sq-type NEAs can collide with the Earth in 0.5 My (although the actual 
probability of such a collision is low). Conversely, none of the remaining 13 NEAs and 7 Mars-crossing
asteroids (MCAs) included in the study, all S and Sq, have any chance of impact in 0.5 My. The time interval of 
0.5 My used in the B10 study was motivated by the recent estimate of the SW timescale in Vernazza et al. (2009) 
where it was proposed, based on a comparative study of asteroid families and OC meteorites, that SW
acts on $<$1~My to fully alter an asteroid spectrum from Q to S. Adopting this timescale, B10 computed 
that the probability that {\it all} known Q-types would fall into the former category by chance (asteroids with 
a possible impact) is only $(75/95)^{20}\approx0$.9\%. Thus, they concluded, the lack of Qs in the latter
group is unlikely to happen by chance. 

The B10 study therefore showed that the Mars-crossing asteroids and distant NEAs on orbits decoupled from the Earth 
(i.e., those that have large $q$ and no Earth encounters at all) are not Qs. This trend has 
already been noted before (Nesvorn\'y et al. 2004, 2005, Marchi et al. 2006). The statistical 
significance assigned to the results by B10 sensitively depends on the selected sample. For example, the significance 
of `not finding Qs among 20 distant NEAs and Mars-crossers' would drop to only about 2$\sigma$ if Mars-crossers were 
excluded from the analysis, as they should because they are not NEAs. If, on the other hand, all non-NEAs were 
included in the analysis, the result becomes obvious because there are no known Q-type asteroids in the 
main belt (except those in the young families). We discuss this issue in more detail in \S3.

(2) B10 used the observed fraction of Q-type NEAs to estimate that encounters up to 16 Earth radii (i.e., $\approx$5 
$R_{\rm Roche}$) should give the asteroid surface a Q-type appearance. This is at odds with our 
understanding of the effects of tidal gravity because it would be surprising if these very distant encounters 
could lead to any displacement of the surface regolith (e.g., Richardson et al. 1998, Walsh and Richardson 2008). 

This problem could indicate that some of the assumptions used in B10 may be invalid. For example, studies of 
asteroid families, lunar craters and some laboratory experiments suggest that the SW timescale, or at least the 
late stage of SW when the regolith gardening processes presumably become important, can last $\gg$1 My (e.g., 
Pieters et al. 2000, Sasaki et al. 2001, Jedicke et al. 2004, NJWI05, Willman et al. 2008). On the other hand, 
experiments with the He ion bombardment of olivine powders conducted in Loeffler et al. (2009) suggest the SW 
timescale $\ll$1 My at 1 AU. The specific choice of the SW timescale in B10 is therefore not very well justified.  
 
Here we study encounters of NEAs to the terrestrial planets and show that the B10's analysis was incomplete. It 
turns out to be important, as originally proposed by NJWI05, to account not only for the encounters of NEAs to Earth but 
also to Venus. Once Venus encounters are included (see \S3), the required encounter distance drops from 16 
to $\approx$10 planetary radii (i.e., by a factor of $(16/10)^3\approx 4$ in the strength of tidal perturbation). In 
addition, we find that the critical distance of planetary encounters, $r^*$, sensitively depends on the the assumed 
SW timescale. When the later is set to 1 My, for example, $r^* \approx$5-7 planetary radii, which is a much more 
reasonable value than the B10 estimate. 

The method used in B10 (see following \S3) has its limitations because it is difficult, even in the statistical sense, to 
reconstruct the history of past planetary encounters by numerical integrations of present orbits into the past. To circumvent
this problem, in \S4 we develop a NEA model by forward orbital integrations of asteroids from their sources in the asteroid 
belt. This method is similar to that used by Bottke et al. (2002) only this time we focus on the statistics of encounters 
of NEAs to the terrestrial planets. The NEA model allows us to consider a wide range of SW timescales, including the 
long ones that cannot be studied by backward integrations of orbits. 

\section{Analysis of Planetary Encounters}

We selected 95 NEAs and Mars-crossing asteroids (MCAs) with known Q, Sq or S taxonomic classification from the near-IR 
taxonomic catalog (DeMeo et al. 2009; Fig. \ref{ics}). This selection is identical to that in B10; see Table 1 in
Supplementary Material of B10 for the complete list. Starting from the current epoch we numerically integrated the
orbits of the selected objects into the past and recorded all planetary encounters in 1 My (longer timescale is
considered in section 4). This encounter record 
needs to be analyzed statistically because the integrated orbits are strongly chaotic.  

In addition to the nominal orbit of each object we also followed 100 orbital clones. The clones were normally 
distributed within the appropriate orbit-determination uncertainty limits around the nominal orbit 
(both taken from NEODyS\footnote{http://newton.dm.unipi.it/neodys/}). In addition to gravitational perturbations 
from 8 planets (Mercury to Neptune), the orbits were also subject to the Yarkovsky force whose strength was chosen 
to sample the full range appropriate for each NEA's size (Bottke et al. 2006). We used the Swift integrator 
(Levison and Duncan 1994) and 1 day time step. We found that shorter time steps produce results that are statistically 
equivalent to those obtained with the 1 day step.

\begin{figure}
\epsscale{0.7}
\plotone{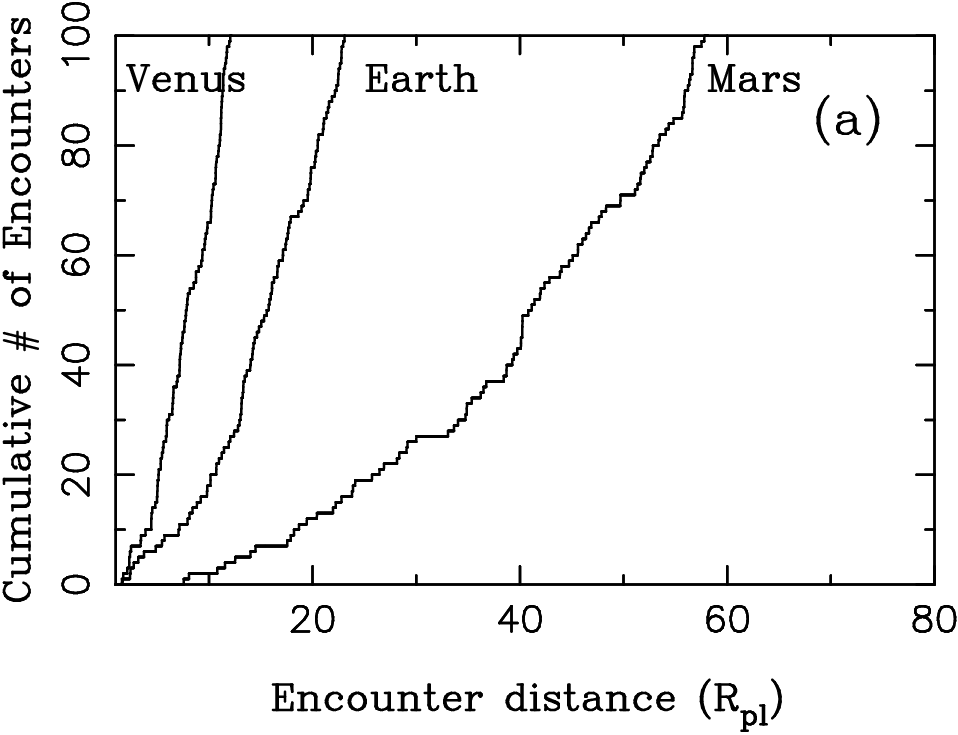}\\[3.mm]
\plotone{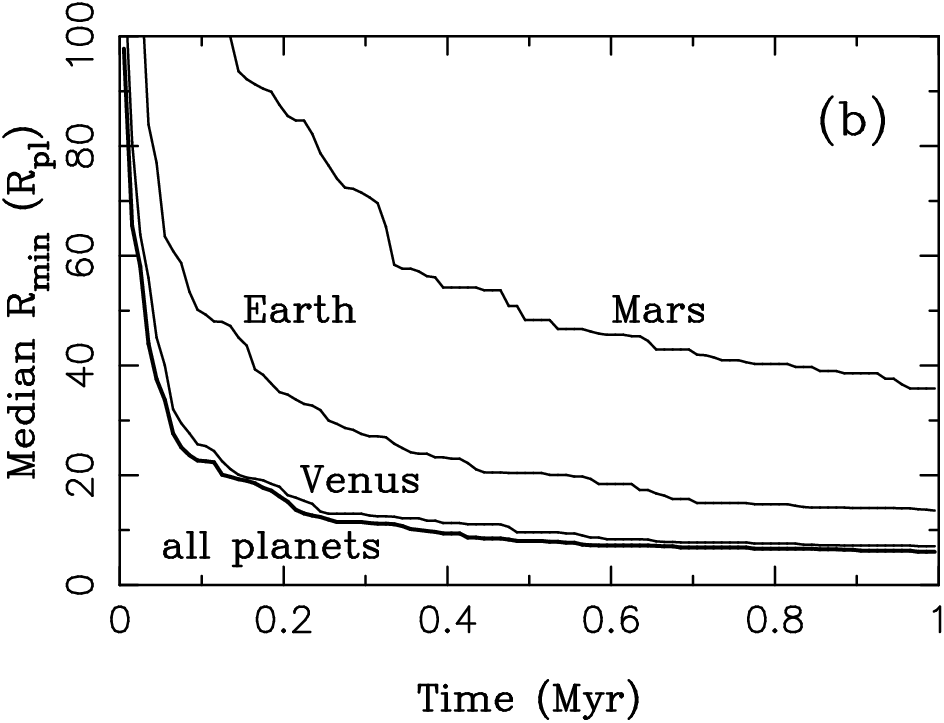}
\caption{The distribution of encounters for $t=0.5$ My (panel a) and $\bar{R}_{\rm min}(t)$ (panel b) 
for asteroid 1862 Apollo. Different lines correspond to encounters with Venus, Earth and Mars.
Close encounters of 1862 Apollo with Mercury are rare. This plot illustrates that the 
encounters with Venus are the dominant type of encounters for NEAs such as 1862 Apollo. The
bold line in (b) shows $\bar{R}_{\rm min}(t)$ when encounters to all planets are considered. In 
(a), a hundred of closest encounters are shown.}
\label{apollo}
\end{figure}

By analyzing planetary encounters recorded in our integrations we found that the encounters with Venus are 
as important as those with the Earth. To show this, we normalized the distance of encounters to each 
planet by the planetary radius, $R_{\rm pl}$, which is 
roughly an appropriate scaling for tidal gravity (e.g., Richardson et al. 
1998), and calculated the minimal encounter distance, $R_{\rm min}(t)$, reached in time $t$. This calculation was 
done for all clones of each individual object. We then determined the median $\bar{R}_{\rm min}$ over clones. 
The median minimum distance has the following statistical meaning: a NEA with given $\bar{R}_{\rm min}(t)=r$ has a 
50\% chance to have close encounter at less then $r$ planetary radii from a planet in time $t$. For illustration, 
Fig. \ref{apollo} shows the distribution of encounters for $t=0.5$ My and $\bar{R}_{\rm min}(t)$ for asteroid 
1862 Apollo. Note, for example, that $\bar{R}_{\rm min}(t=0.5{\rm My})$ for the encounters of 1862 Apollo to 
Venus and Earth are $9.5$ and $20.3$ $R_{\rm pl}$, respectively, while it is only 8 $R_{\rm pl}$ when all 
planetary encounters are combined. 

\begin{figure}
\epsscale{0.7}
\plotone{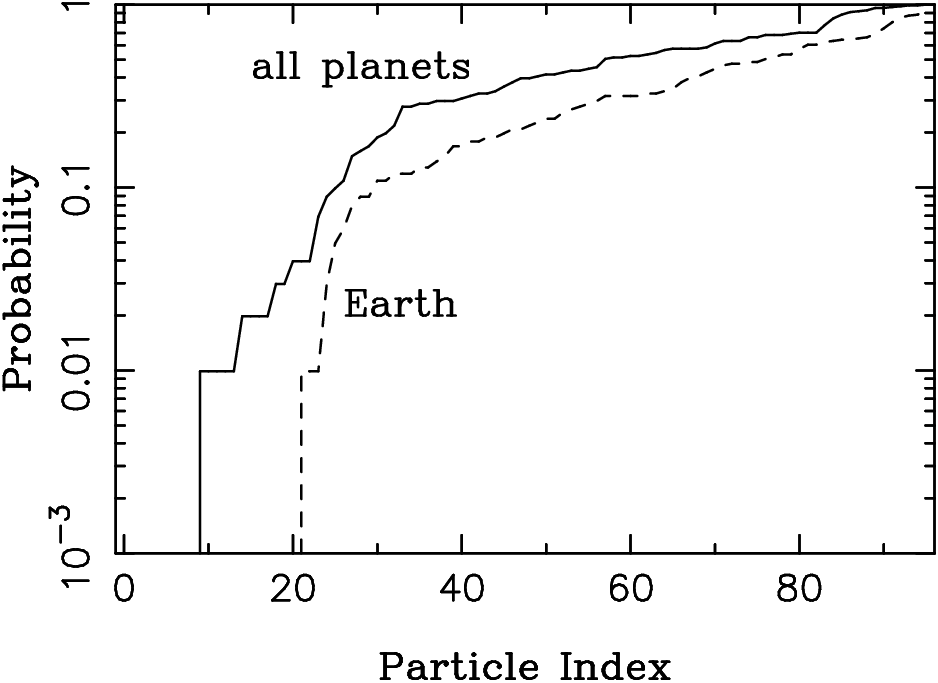}
\caption{The probability of having an encounter with $r<20$ $R_{\rm pl}$ in $t=0.5$ My. Particle
index denotes individual objects that were ordered by the increasing probability value. The dashed and 
solid lines show the probability for encounters with the Earth and all planets, respectively.} 
\label{dprob}
\end{figure}

In the next step, we searched for objects among the 95 NEAs and MCAs included in this study that have a 
negligible probability of having a close encounter with any planet. To quantify this, we calculated the probability 
$P(R,t)$ that an individual object in our sample had an encounter with $r<R=20$ $R_{\rm pl}$ in the past 
$t=0.5$ My (Fig. \ref{dprob}). We found that nineteen out of twenty S/Sq asteroids listed in B10 as having 
MOID\footnote{The Minimum Orbital Intersection Distance or MOID is defined in Bonanno (2000). It is a 
useful indication of whether or not two objects can collide and is frequently used to identify the potentially 
hazardous asteroids. The information carried in MOID, however, does not indicate whether such an collision 
(or close encounter) is likely or not; that depends on the exact location of the two objects in their orbits. 
Using MOID, B10 divided objects into those with MOID corresponding to Earth encounters smaller than the 
lunar distance and those with MOID larger than this distance.} outside the lunar distance also have 
$P(20,0.5)<5$\%; only 54690 2001EB has $P(20,0.5)=10$\% of having encounter with Mars. 

What is slightly more puzzling is that three asteroids with B10's MOID values in the lunar distance also have 
$P(20,0.5)<5$\%. One of these is classified as S (719 Albert), and two are Qs (162058 1997AE12 and 2008CL1).
This shows that the classification of objects based on MOID is ambiguous because it does not properly take 
into account the actual encounter probability over a finite time interval. It is therefore incorrect to assign the 
B10's result (even approximate) statistical significance, because such a calculation will depend on the subjective 
choice of the cutoff value. For example, the partition of Q-type objects between $P(20,0.5)>5$\% and $P(20,0.5)<5$\% is 
not statistically unusual (unless MBAs were taken into account).\footnote{So far there is not known any Q-type 
object among distant MCAs with $P(20,0.5)<1$\% (for which we have the near-IR data). It will be interesting 
to see if this situation holds with new observations.}

When only encounters to the Earth are considered, the classification of objects based on their encounter 
probability becomes less ambiguous (Fig. \ref{dprob}). This happens because the probability of Earth encounter is 
a step-like function with either $P(20,0.5)>10$\% or $P(20,0.5)<1$\%, and very few objects (6 in total; 3288, 
5143, 6047, 23187, 2006NM and 2001FA1) in the intermediate range. One of these six intermediate objects, 5143 Heracles, 
is a Q with an Earth-encounter probability $P(20,0.5)=9$\%. Our 20 objects with $P(20,0.5)<1$\% also have large 
MOID for Earth encounters according to B10. 

We will consider two cases in the following text. In the first case, we will assume that the main effect 
on surface regolith of an asteroid is driven by tidal gravity during the asteroid's encounters to 
the terrestrial planets (case 1). All planets, mainly Venus and Earth, must be considered in this case. To 
compare our models with the data, all 95 objects with known near-IR taxonomy will be considered as one group. 
In the second case, we will consider encounters to the Earth only (case 2). There is a possibility (discussed 
in more detail in \S5) that electrically charged regolith particles (e.g., by photoelectric effect; Lee 1996) 
can be lofted by the Lorentz force when the asteroid passes through the Earth's magnetosphere. Since the 
Earth magnetosphere extends to a larger distance than where tidal gravity could be important, its effects may 
potentially be relevant for distant encounters. Distant encounters with Venus and Mars need not to be 
considered because these planets do not have important magnetic fields. In this case, we will discard 
20 objects with $P(20,0.5)<1$\% for Earth encounters (group 2 in the following) from our list and 
consider the remaining 75 objects only (group~1).      

The SW timescale and critical encounter distance for which the tidal gravity (or Lorentz force) can be important 
are treated as free parameters in the following. Specifically, we determine the number of bodies in the selected 
sample that are expected to have at least one encounter with distance $r<r^*$ in time $t$, where $t$ and $r^*$ are 
free parameters. This value gives us a sense of the expected fraction of the Q-type objects in our model
as a function of the SW timescale and $r^*$.\footnote{Note that both the tidal gravity and Earth's magnetospheric 
effects should strongly decay with the encounter distance. It is therefore approximately correct to assume 
that the surface is reset if a close approach is made within $r^*$ and otherwise unaffected. A more realistic 
resurfacing model would include more free parameters and would be difficult to constrain with the present data.}
As in B10, we use a definition of the SW timescale, $t_{\rm sw}$, 
as the characteristic time interval during which an initially `fresh' Q-type NEA affected by SW {\it remains Q}. 
This is the natural timescale that is directly constrained by the observed Q-type fraction among NEAs.

Note that our definition differs from the one used elsewhere (e.g., Jedicke et al. 2004, NJWI05, Willman et al. 2008,
Vernazza et al. 2009), where the SW timescale was defined as the time interval for SW to approach/reach completion. 
Additional assumptions on the SW dependence on time are therefore required to compare $t_{\rm sw}$, as determined 
here, with the SW timescales estimated elsewhere. For example, studies of asteroid families suggest that SW can  
partially weather a surface in $\sim$1 My (Chapman et al. 2007, Moth\'e-Diniz and Nesvorn\'y 2008, Vernazza et al. 
2009), and then proceed towards completion during a phase that can last several Gys (Willman et al. 2008).
The timescale $t_{\rm sw}$ that we determine in this work provides constraints on the initial stages of the 
SW process. 

\begin{figure}
\epsscale{0.485}
\plotone{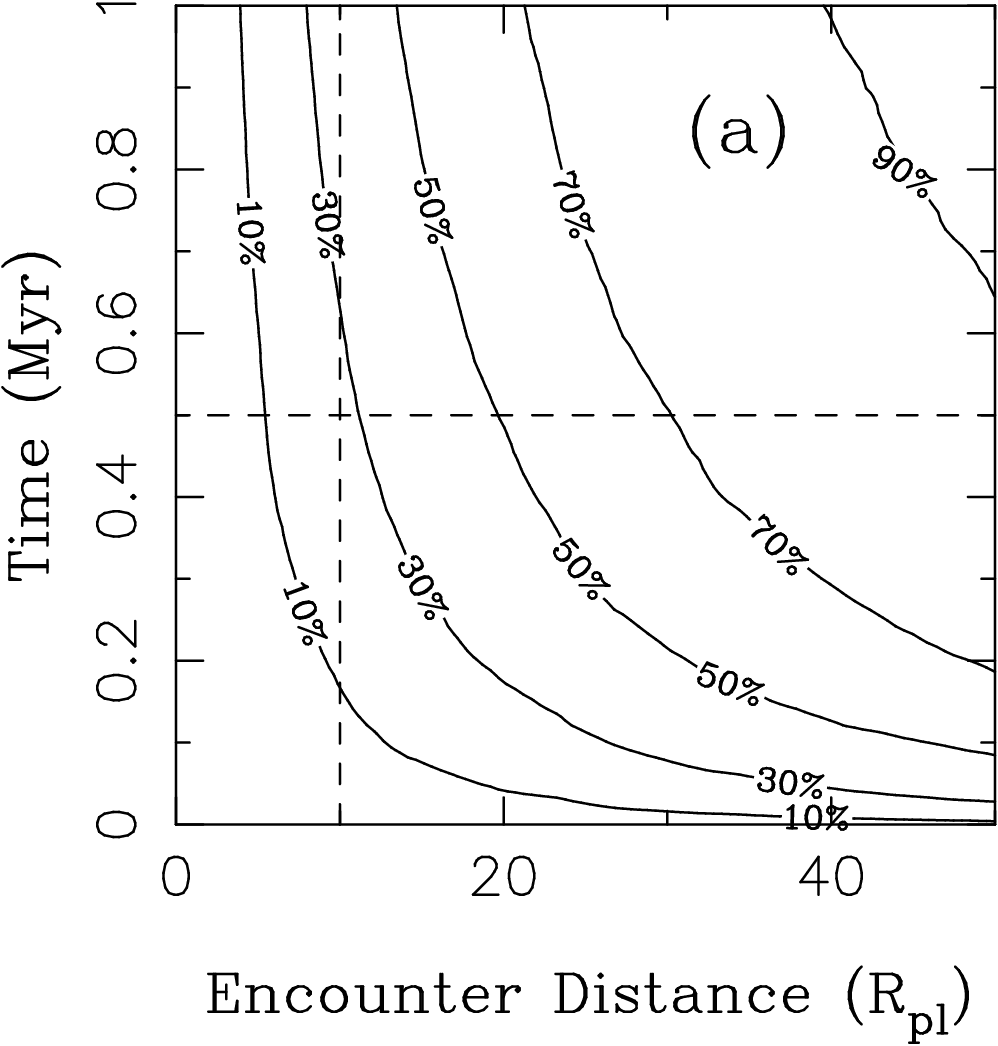}\hspace*{5.mm}
\epsscale{0.495}
\plotone{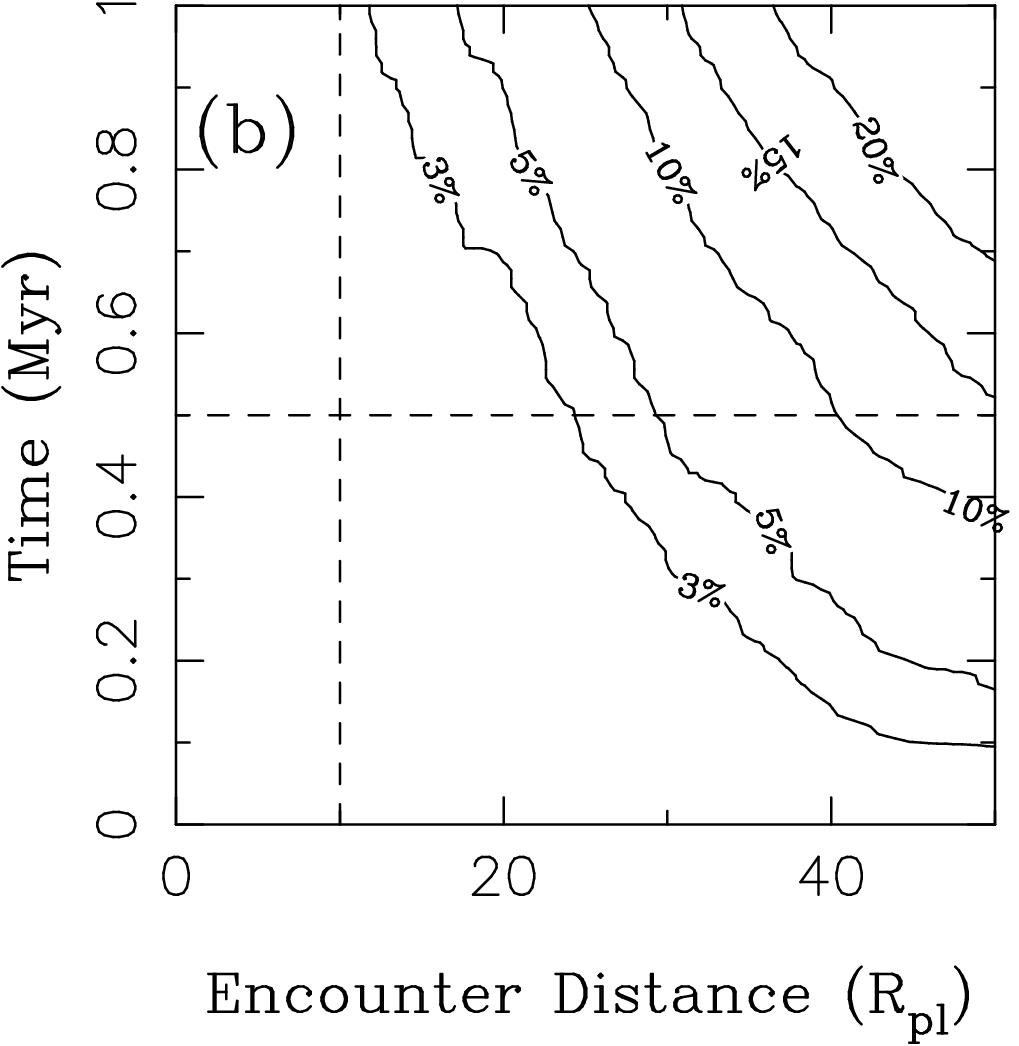}
\caption{The fraction of objects having a planetary encounter with distance $r<r^*$ in time $t$: (a) group 
1 (75 NEAs with high encounter probability); and (b) group 2 (20 NEAs and MCAs with low encounter
probability). All planets are considered here. For reference, the dashed lines show $t=0.5$ My and 
$r^*=10$~$R_{\rm pl}$. With $t=0.5$ My we find that roughly 28\% of objects in group 1 have a planetary 
encounter with $r<10$ 
$R_{\rm pl}$. We do not have enough resolution in the model to directly compute the fraction of group 2 
objects for $t=0.5$ My and $r^*=10$ $R_{\rm pl}$. By extrapolating the trend from larger $t$ and $r^*$, 
we roughly estimate that this fraction is $\sim$0.5\%.}
\label{prob1}
\end{figure}

Figure \ref{prob1} shows the expected fraction of Q-type objects in case 1, as defined above, as a function of 
$t_{\rm sw} \le 1$ My and $r^*$ (see \S4 for $t_{\rm sw} > 1$ My). 
We find that $<$1\% of group 2 objects have planetary encounter with $r<r^*=10$ $R_{\rm pl}$ in $t=0.5$~My. This 
fraction increases to nearly 3\% for $t=1$ My. Since there are only 20 objects in group 2, it is therefore 
statistically unlikely that one (or more) object(s) in group~2 would have a recent encounter with $r<r^*=10$ 
$R_{\rm pl}$. This is consistent with current observations that indicate that none of these objects is a 
Q. Spectrophotometric observations of at least $\sim$100 group-2 objects would be needed to test the NJWI05
model in a more stringent way. 

\begin{figure}
\epsscale{0.7}
\plotone{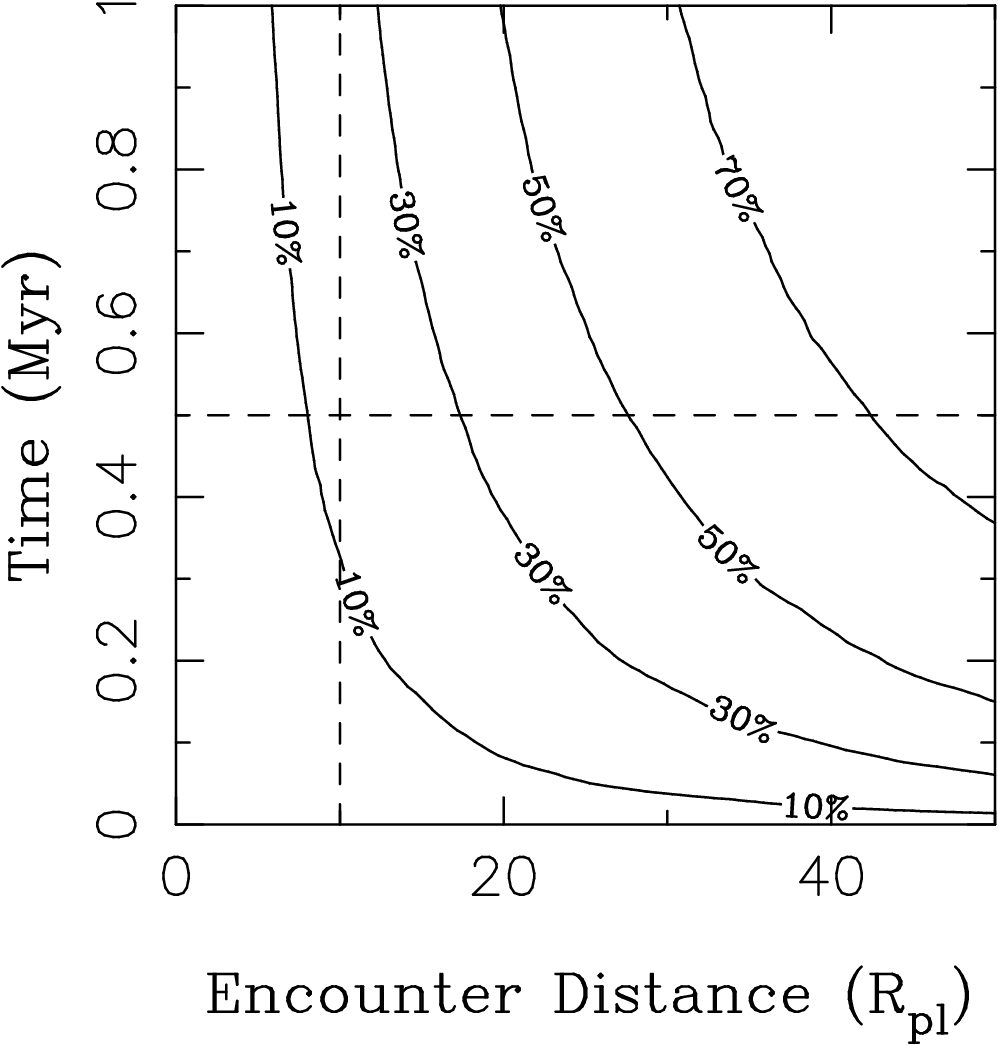}
\caption{The fraction of group-1 objects having Earth encounter with distance $r<r^*$ in time $t$.
With $t=0.5$ My we find that roughly 28\% of objects in group 1 have Earth encounter with $r<17$ 
$R_{\rm Earth}$. For reference, the dashed lines show $t=0.5$ My and $r^*=10$~$R_{\rm pl}$.} 
\label{prob2}
\end{figure}

Group-1 NEAs are those that have a large number of encounters with the terrestrial planets, mainly Venus and Earth. 
These two planets are equally important. For example, B10 estimated by neglecting Venus encounters that 
Earth encounters of group-1 NEAs with $r^*=16$ $R_{\rm Earth}$ are needed, if $t_{\rm sw}=0.5$ My, to explain 
the observed fraction of Q types (28\%, see below). Here we repeat this calculation and find $r^*=17$ 
$R_{\rm Earth}$ (Fig. \ref{prob2}), a slightly larger value than the B10 estimate but in a reasonable agreement 
with it (the difference can be explained by our larger statistics). Now, including Venus encounters but neglecting 
those to the Earth we find that the Venus encounters with $r^*=19$ $R_{\rm Venus}$ would be required. When both the 
encounters to Venus and Earth are considered, however, the required encounter distance drops to 
$r^*=10$~$R_{\rm pl}$ (Fig. \ref{prob1}a). This casts doubt on the claims in B10, where it was suggested 
that the Earth's tidal gravity can reset the NEA surface during encounters at 16~$R_{\rm Earth}$.

\begin{figure}
\epsscale{0.7}
\plotone{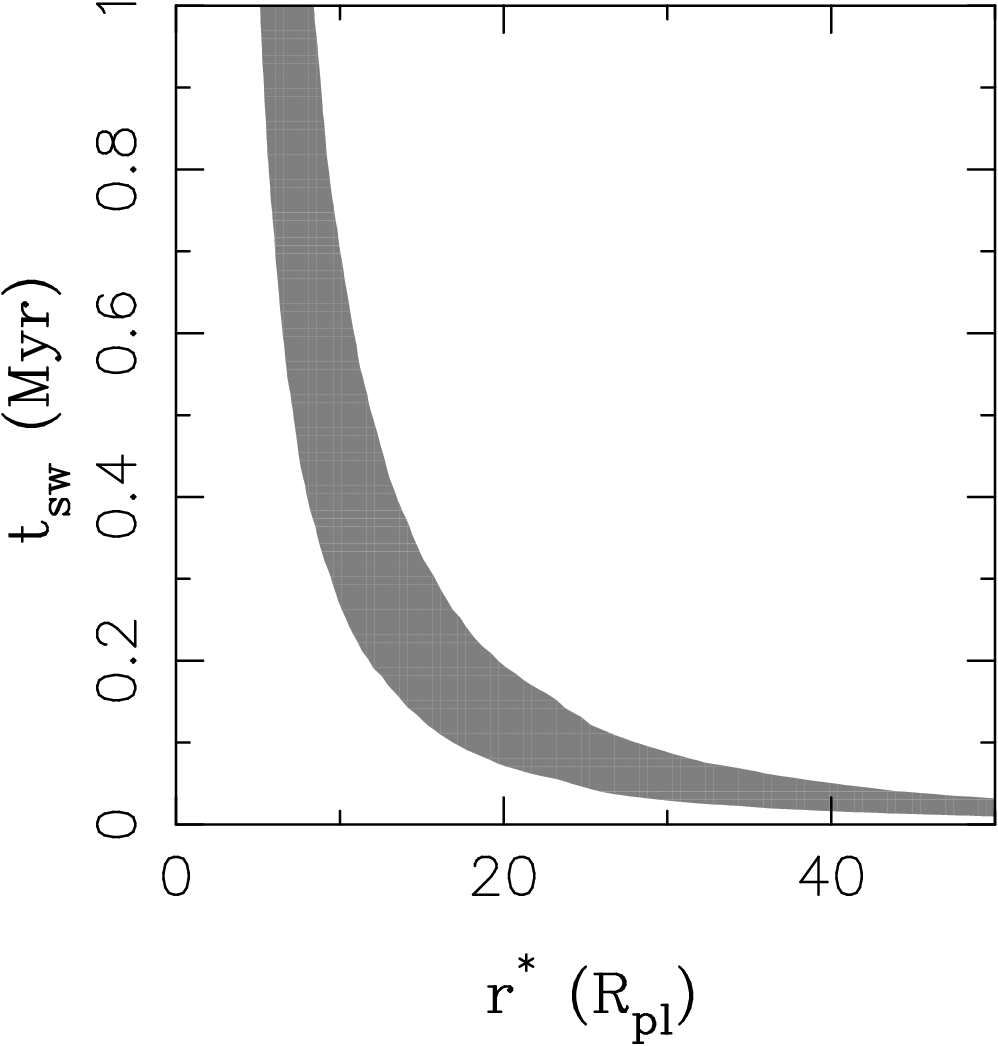}
\caption{The preferred solutions according to the model described in \S3. The shaded region shows the 
range of fractional values between 16\% to 26\%. All 95 selected asteroids and encounters to all
planets were considered here (case 1). The range of preferred solutions remains nearly the same when only
group-1 NEAs are considered. This is because their slightly larger overall encounter probability 
is compensated by a larger fraction of Q-type NEAs observed in that group.}
\label{prob3}
\end{figure}

The observed fraction of Q-type NEAs can be used to constrain parameters $r^*$ and $t_{\rm sw}$ (Fig. \ref{prob3}). 
This fraction is $20/95=0.21$ when all objects are considered (case 1 as defined above) and $20/75=0.28$ when only 
objects in group 1 are considered (case 2). In either case, the best-fit solutions are located along a hyperbola-shaped 
region in $(r^*,t_{\rm sw})$ space. Since planetary tides during encounters with $r>r^*=20$ $R_{\rm pl}$ should be 
negligible, we find that $t_{\rm sw}>0.1$ My. This result holds unless the Earth's magnetospheric effects 
are important at $r>35$ $R_{\rm Earth}$ (Fig. \ref{prob2}), which is unlikely.

We find that $r^*\approx8$-12 $R_{\rm pl}$ for $t_{\rm sw}=0.5$ My. This $r^*$ value is probably too large
compared to the expectations from the simulations of tidal effects during planetary encounters (Richardson et al. 
1998, Walsh and Richardson 2008). These simulations show that the large-scale effects of tidal gravity 
should be minimal beyond $\sim$6 $R_{\rm pl}$ even in the most favorable case of fast `prograde' rotation of the 
small object. Here, the prograde rotation is defined with respect to the encounter trajectory. We thus believe 
that $t_{\rm sw} \sim 1$ My can probably better fit the available constraints (from NEAs and Chapman et al. 2007, 
Moth\'e-Diniz and Nesvorn\'y 2008, Vernazza et al. 2009) because this slightly longer timescale leads to 
$r^*\approx 5$-7 $R_{\rm pl}$. Note that these $r^*$ values are plausible because the optically-active thin 
surface layer may be vulnerable to even tiniest tidal perturbations that were not considered in the simulations 
of Richardson et al. (1998), and Walsh and Richardson (2008). Values $t_{\rm sw}>1$ My are also plausible (based 
on the NEA constraint only) but we are not able to deal with these longer timescales with the method described 
in this section. 

\section{NEA Model}

The method described in the previous section is only approximate because it is difficult, even in the 
statistical sense, to reconstruct the history of past planetary encounters by numerical integrations of 
present orbits into the past. It is even more problematic to try to extend these numerical integrations 
beyond 1 My, to times comparable with the average orbital lifetime of NEAs ($\approx$5 My; Bottke et al. 2002). 
This is because the statistical results obtained from these integrations cannot be used to retrace the 
real orbital evolution of individual objects from their source locations in the main belt to NEA space. 
Consequently, the encounter statistic obtained from such integrations would be incorrect. A different 
method needs to be used to circumvent this problem (and check on the results obtained in the previous 
section). 

We used the method developed in Bottke et al. (2002; hereafter B02). B02 constructed the NEA model by tracking 
orbits originating from various locations in the main belt, such as the $\nu_6$ and 3:1 resonances, and the 
population known as the Intermediate source Mars Crossers (IMCs for short). IMCs have marginally unstable orbits 
that are leaking from more stable locations in the inner main belt but have not yet reached Mars-crossing 
space (Migliorini et al. 1998, Morbidelli and Nesvorn\'y 1999). By calibrating the orbital distribution
obtained in the model to that of known NEAs, B02 was able to set constraints on the contribution of each 
source to the NEA population as a function of absolute magnitude $H$. Apparently, the three most important sources 
are the $\nu_6$ resonance, 3:1 resonance and IMCs, which contribute by 37\%, 20\% and 27\%, respectively,
for $H<18$. [The outer main belt resonances and Jupiter-family comets provide the remaining 16\%.]

We conducted numerical simulations similar to those reported in B02 only this time focusing on the statistics 
of close encounters of NEAs with the terrestrial planets. Specifically, we tracked orbits of $\sim$1000 test 
particles (per source) as they evolve from their source regions into planet-crossing space. These integrations 
included seven planets (Venus to Neptune). Thermal effects on orbits (such as the Yarkovsky effect) 
were neglected because NEA dynamics is mainly controlled by planetary encounters and powerful resonances. We 
used a variant of the Wisdom-Holman map (Wisdom and Holman 1991) known as Swift\_rmvs3 (Levison and Duncan
1994). We modified the Swift integrator so that it records all encounters of model NEAs with planets up to 
a distance of 20 $R_{\rm pl}$. 

These data were used in a statistical model that follows the orbital evolution of each object and estimates 
its spectral index at any given moment. We define the spectral index, $I_s$, as 0 for a fresh Q-type 
object and 1 for a fully space weathered S type object. The intermediate values $1/3<I_s<2/3$ are used 
to represent the Sq-type asteroids. The model has two parameters: $r^*$ and $t_{\rm sw}$. Each object
is assumed to be initially fully space weathered with $I_s = 1$. If an encounter with $r<r^*$ 
occurs, we set $I_s = 0$ at the corresponding time, and let a simple SW algorithm increase $I_s$. Therefore, 
assuming that no additional encounters with $r<r^*$ happen in the interim interval, $I_s=1/3$ after 
$t_{\rm sw}$ has elapsed. Parameter $t_{\rm sw}$ thus represents the timescale during which an initially 
fresh Q-type asteroids remains Q. This definition is consistent with the one used in the previous section.

Note that our algorithm is only a simple representation of the SW process that, in reality, must be more 
complicated. For example, Jedicke et al. (2004) and Willman et al. (2008) assumed that the spectral slope has 
an exponential dependence on time (as if SW were produced by constant SW agent), and defined the SW timescale 
as the characteristic exponential time scale, $\tau$, of this dependence. The relationship between our $t_{\rm sw}$ 
and their $\tau$ is $t_{\rm sw} = -\ln(2/3) \tau \approx 0.4 \tau$, where $\tau \sim 1$ Gy in Willman et al. (2008).
This suggests that $t_{\rm sw}$ could be very long. On the other hand, Vernazza et al. (2009) invoked a two-step 
process with the fast initial stage, perhaps due to ion sputtering, and slower later stage, as in Willman et al. 
(2008). These two-step process would indicate that $t_{\rm sw}\lesssim1$ My (Chapman et al. 2007, Moth\'e-Diniz and 
Nesvorn\'y 2008).

We run our code over all test orbits and record $I_s$ as a function of $a$, $e$ and $i$. The
expected fraction of Q-type NEAs in a given orbital bin, $f_Q(a,e,i)$, is then estimated as 
$f_Q=N(I_s<1/3)/N$, where $N$ and $N(I_s<1/3)$ are the total number of recorded cases and the number 
of cases with $I_s<1/3$, respectively. The contribution of particles starting in different sources is 
weighted by the relative importance of each source according to B02. Fractions $f_Q(a,e,i)$ 
obtained in this NEA model with different $r^*$ and $t_{\rm sw}$ are then compared with the observed 
fraction of Q-type NEAs. This comparison helps us to set constraints on the SW timescale and 
critical encounter distance.

\begin{figure}
\epsscale{0.7}
\plotone{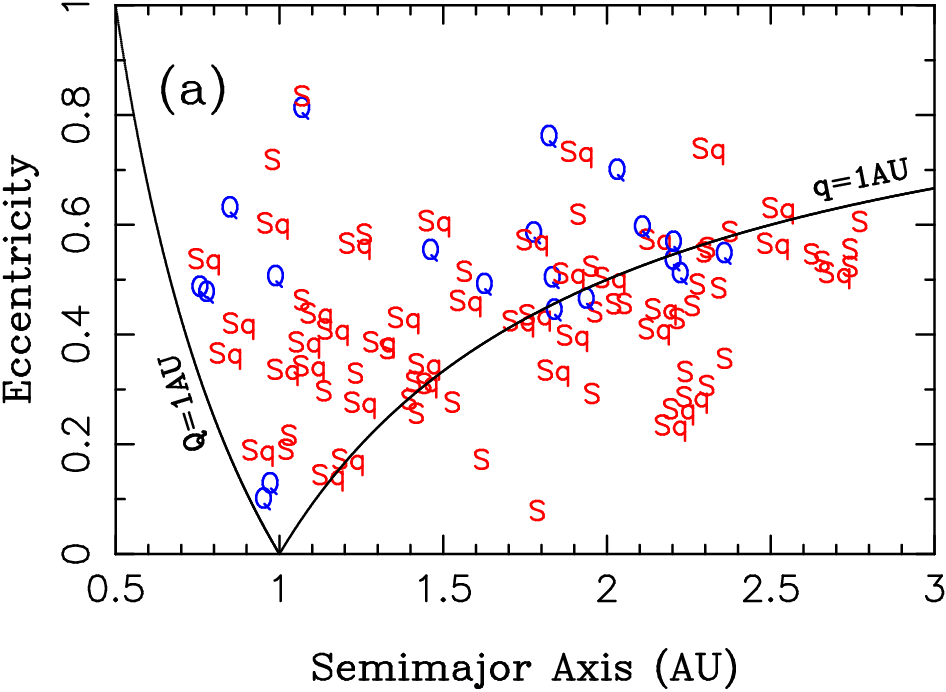}\\[5.mm]
\plotone{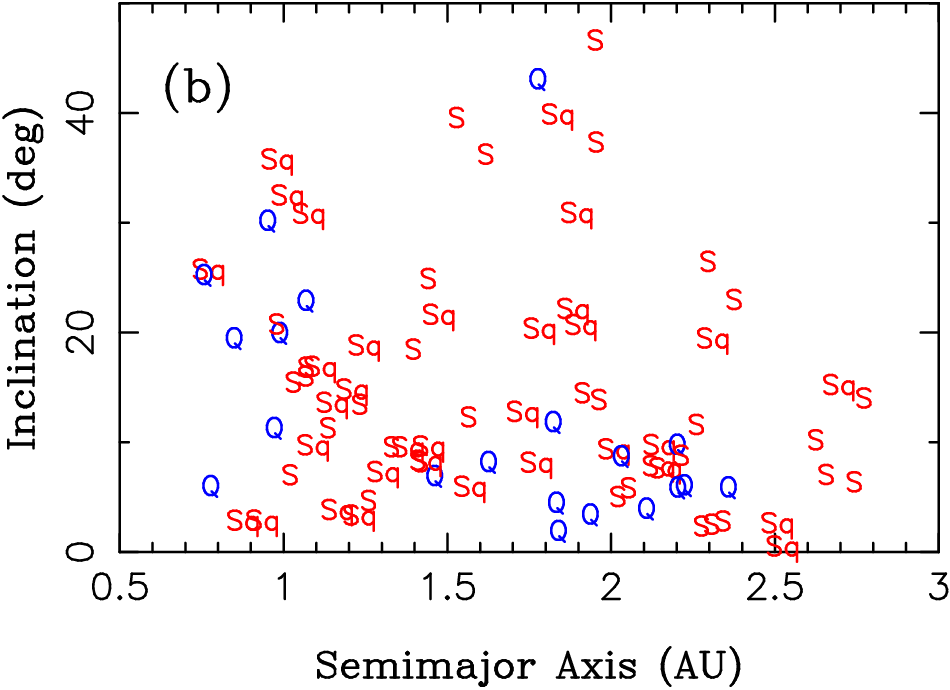}
\caption{The proper orbital elements of 95 NEAs and MCAs included in this work: (a) proper semimajor axes and 
proper eccentricities, and (b) proper semimajor axes and proper inclinations. The proper elements of individual 
orbits were calculated as the arithmetic mean of minimum and maximum values given by Gronchi and Milani (2001). 
The symbols show the taxonomic classification of objects based on visible and near-infrared spectra (DeMeo 
et al. 2009). The solid lines in (a) show the proper perihelion and proper aphelion distances of 1 AU.}
\label{ics}
\end{figure}

With only 20 known Q-type objects the current spectrophotometric catalog of NEAs is largely incomplete 
and probably biased by the observer's selection criteria that are difficult to characterize. We do 
not make any attempt to compensate for the observational bias. To compare our model with the sparse 
data, we find that the best strategy is to divide the orbital region into large bins in the perihelion 
and aphelion distance, and inclination. This is useful because large bins allow for better statistics. 
It is also better to use $q$, rather than $a$ or $e$, because the orbital distribution of Q-type
NEAs has a sharp edge at $q=1$ AU with no Q-types known with $q>1.1$ AU (Fig. \ref{ics}). 

Simulated fraction $f_Q$ is compared with observations using the usual $\chi^2$ statistics. Given the dependence 
of the statistics on the bin selection, however, we do not attempt to assign any formal confidence levels to various 
parameter choices. Instead, we only compare different models relatively among themselves according to 
their $\chi^2$ value; models with smallest $\chi^2$ are given priority.

\begin{figure}
\epsscale{0.485}
\plotone{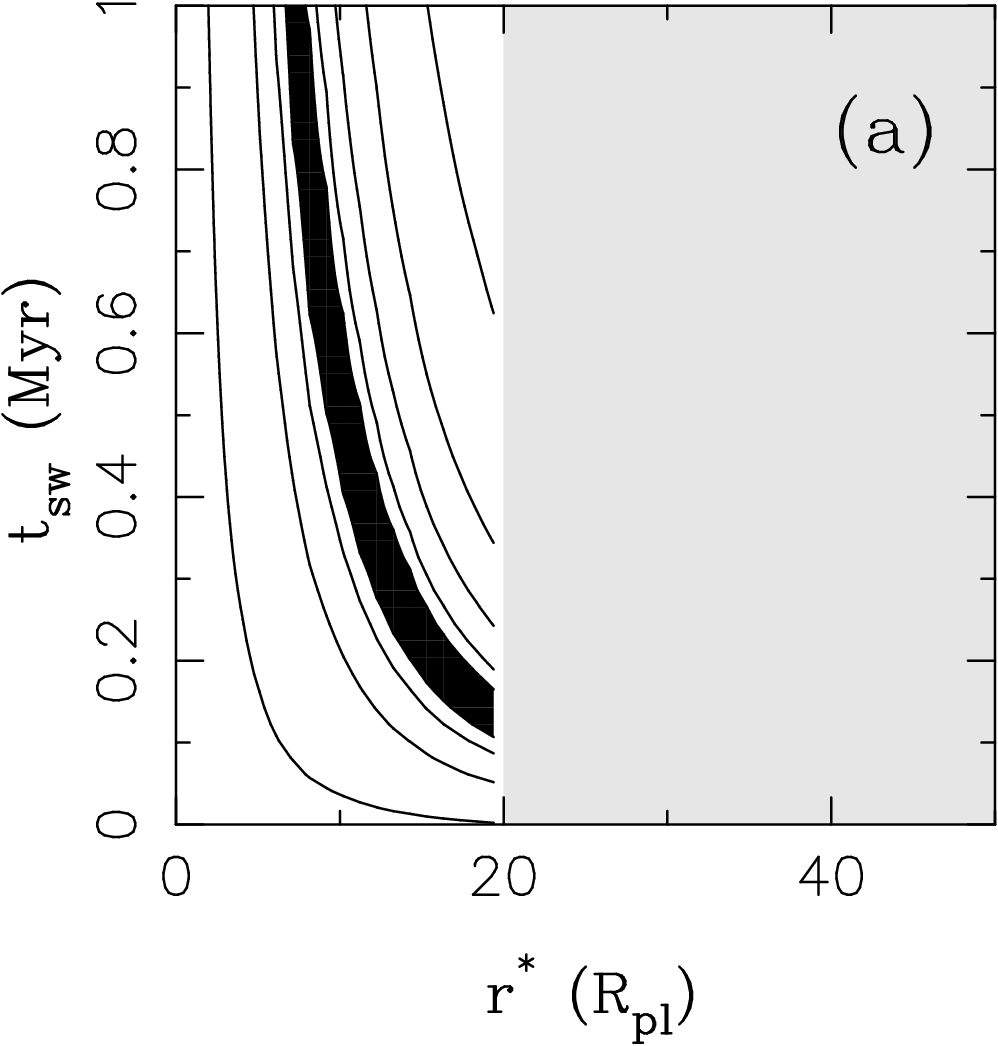}\hspace*{5.mm}
\epsscale{0.5}
\plotone{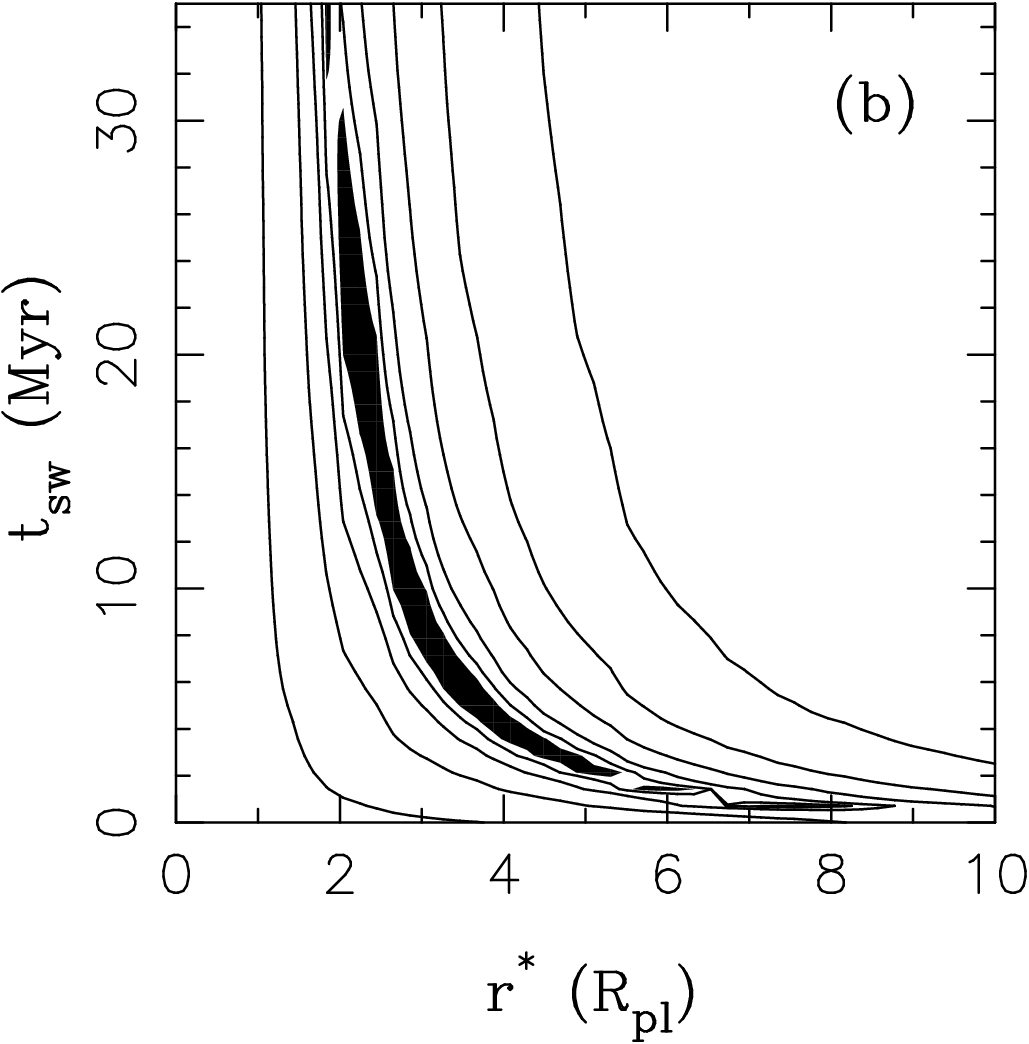}
\caption{The $\chi^2$ contours in ($r^*$,$t_{\rm sw}$) plane. We used the NEA model to estimate
$f_Q(a,e,i)$ for: (a) $t_{\rm sw}<1$ My, and (b) $t_{\rm sw}<35$ My. The distribution was compared 
to observations. The black-shaded area shows our best-fit parameter values. Each pair of lines that 
envelops the shaded area shows $\chi^2$ contours that are spaced by a multiplication factor of 3.
We have not followed encounters with $r>20$ $R_{\rm pl}$ in the NEA model described in \S4. The 
range on X and Y axes in (a) is set to facilitate a direct comparison with Fig. \ref{prob3}.}
\label{fit}
\end{figure}

Figure \ref{fit} shows the $\chi^2$ values for models with different $r^*$ and $t_{\rm sw}$. The range of 
parameter values that fits observations best roughly overlaps with the region identified from backward 
numerical integrations in \S3 (cf. Fig. \ref{prob3}). This gives some credibility to the method used in \S3. 

We find that $t_{\rm sw}<0.1$ My can be rejected unless $r^*>20$ $R_{\rm pl}$, in a good agreement with the 
results obtained in \S3. Figure \ref{fit}b extends these results to $t_{\rm sw}=35$~My. The best fits occur
along a curve that indicates progressively smaller $r^*$ values for longer $t_{\rm sw}$. Eventually, 
the fits following this curve slightly degrade for $t_{\rm sw}>30$ My. Also, $r^*<2$ $R_{\rm pl}$ for 
$t_{\rm sw}>35$~My, while $r^*>2$ $R_{\rm pl}$ according to Richardson et al. (1998). These long
SW timescales therefore do not appear plausible.  

\begin{figure}
\epsscale{0.8}
\plotone{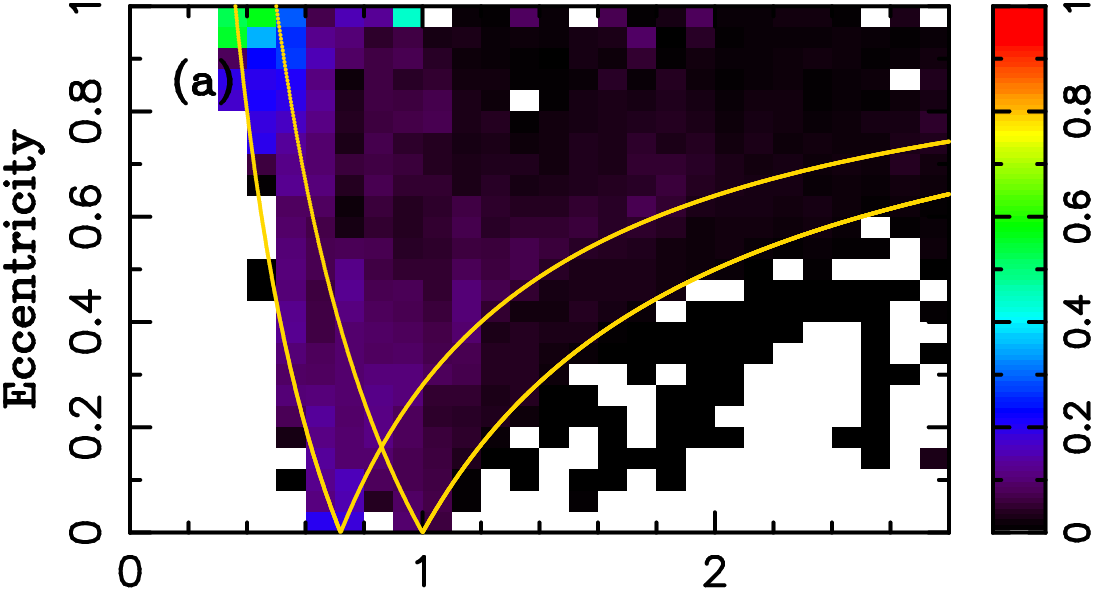}\\[5.mm]
\plotone{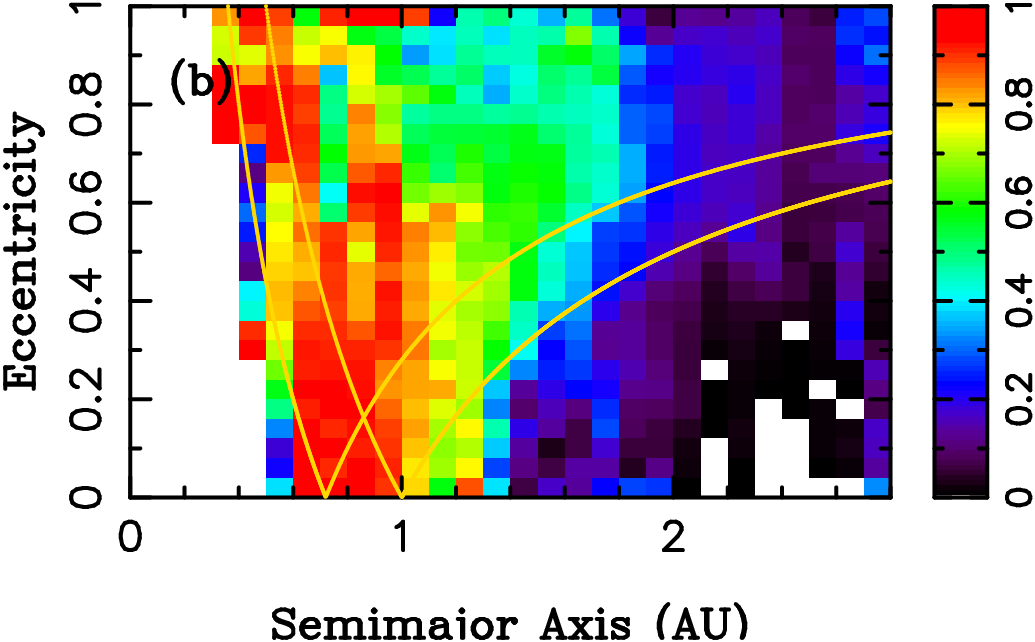}
\caption{The orbital distribution of Q-type objects expected from the NEA model with: (a) $r^*=10$ $R_{\rm pl}$ 
and $t_{\rm sw}=0.1$ My, and (b) $r^*=5$ $R_{\rm pl}$ and $t_{\rm sw}=15$ My. The color scheme shows
the expected number of Q-type objects in each bin as a fraction, $f_Q$, of the total number of objects in 
that bin. The red region shows orbits with the largest model concentrations of Q-type NEAs. Both these model 
parameters are clearly implausible because $f_Q$ is too small in (a) and too large in (b).}
\label{m1}
\end{figure}

\begin{figure}
\epsscale{0.8}
\plotone{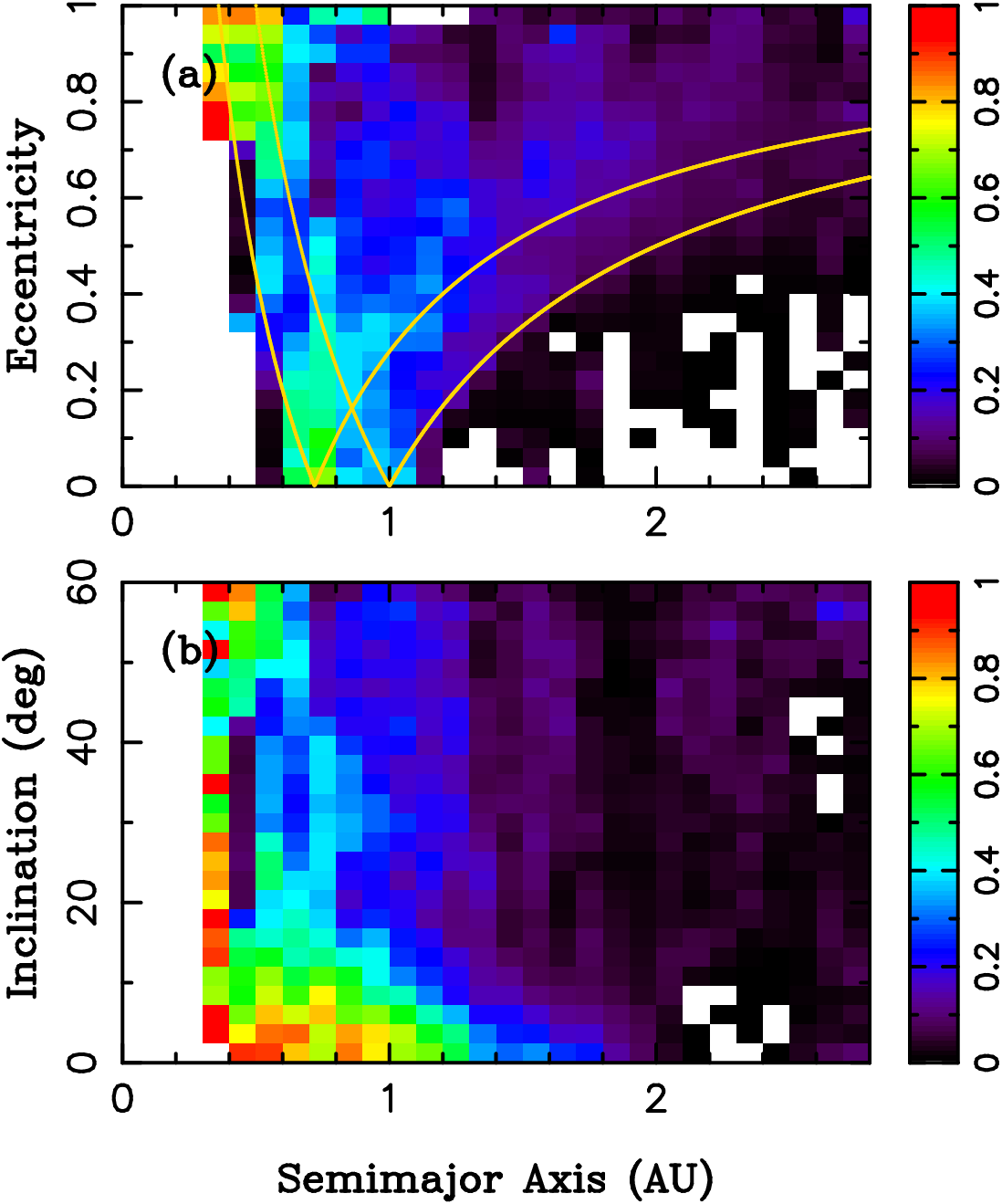}
\caption{The orbital distribution of Q-type objects expected from the NEA model with $r^*=7$ 
$R_{\rm pl}$ and $t_{\rm sw}=1$ My. The color scheme shows the number of Q-type objects in each bin as 
a fraction of the total number of objects in that bin.}
\label{m2}
\end{figure}

\begin{figure}
\epsscale{0.8}
\plotone{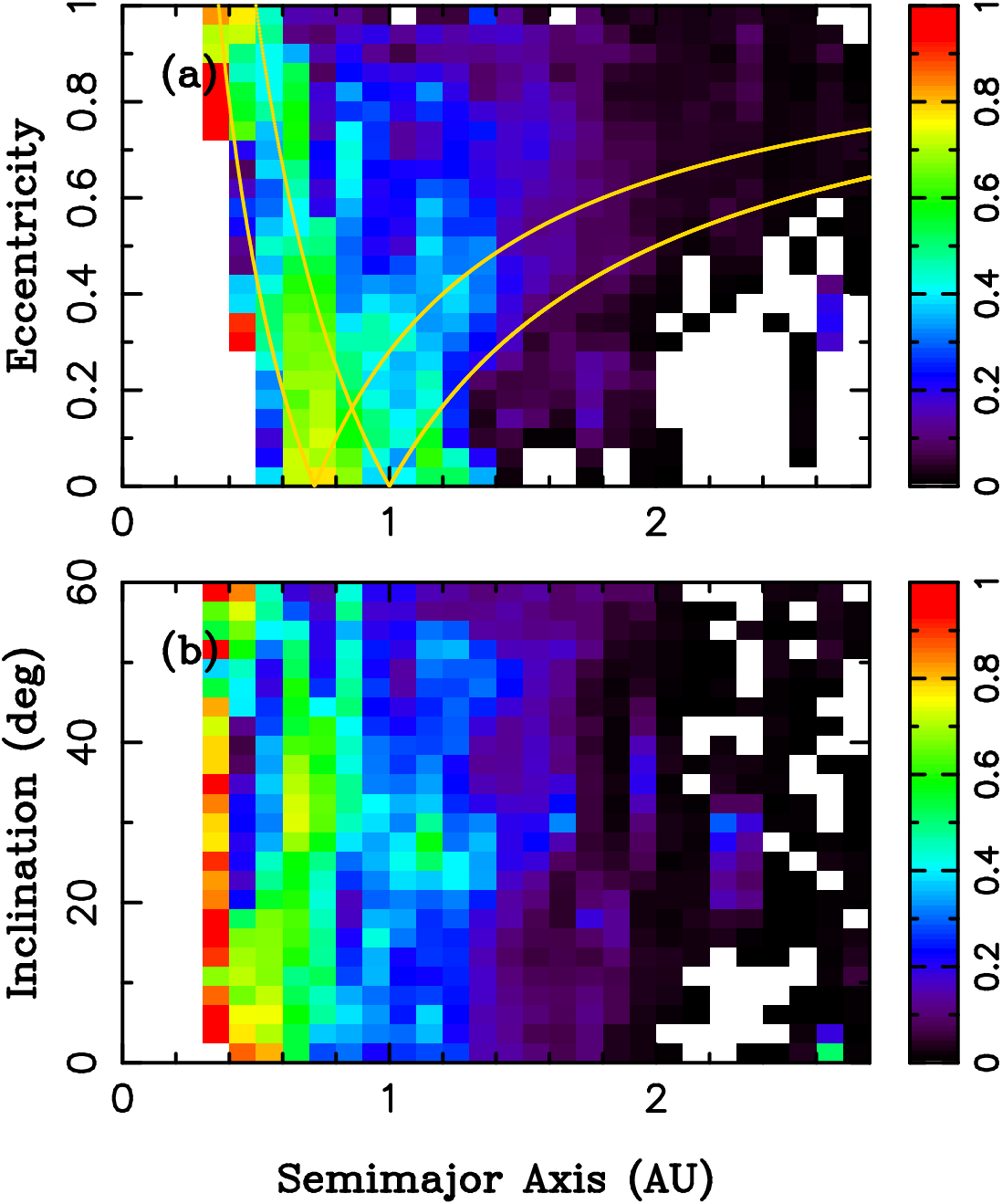}
\caption{The orbital distribution of Q-type objects expected from the NEA model with 
$r^*=2.5$ $R_{\rm pl}$ and $t_{\rm sw}=15$ My. The color scheme shows the number of 
Q-type objects in each bin as a fraction of the total number of objects in that bin.}
\label{m3}
\end{figure}

Fraction $f_Q(a,e,i)$ obtained in our model is shown in Figs. \ref{m1}-\ref{m3} for several different values 
of $t_{\rm sw}$ and $r^*$. Figure \ref{m1} shows $f_Q$ for $r^*=10$ $R_{\rm pl}$ and $t_{\rm sw}=0.1$ My, and 
$r^*=5$ $R_{\rm pl}$ and $t_{\rm sw}=15$ My. Both these parameter choices do not fit observations well. The one
with $t_{\rm sw}=0.1$ My produces an overall excess of Q-type objects with $f_Q>0.5$ for $q<1$ AU and 
$a<2$ AU. The one with $t_{\rm sw}=15$ My shows $f_Q<0.1$. In comparison, the surveyed NEAs have 
$f_Q=0.2$-0.3 overall. Note that the two models illustrated in Fig. \ref{m1} lay outside the low-$\chi^2$
region shown in Fig. \ref{fit}.

Two of our models that match observations better are illustrated in Fig. \ref{m2} ($r^*=7$ $R_{\rm pl}$ and 
$t_{\rm sw}=1$~My) 
and Fig. \ref{m3} ($r^*=2.5$ $R_{\rm pl}$ and $t_{\rm sw}=15$ My). These models correspond to some of the 
lowest $\chi^2$ values that we have obtained. Fraction $f_Q$ increases in both these models with decreasing 
heliocentric distance. In Fig. \ref{m2}, the model with $t_{\rm sw}=1$ My produces a concentration of Q-type
objects with low orbital inclinations, while in Fig. \ref{m3} the model with $t_{\rm sw}=15$ My shows a more 
equal distribution of $f_Q$ in inclination. These differences could be used to discriminate between 
short and long SW timescales, even without an explicit constraint on $r^*$, when spectroscopic observations of 
NEAs become more complete. 

While the overall fraction of Q-type NEAs in Figs. \ref{m2} and \ref{m3} closely matches current 
observations, the model distribution of Qs in orbital space differs in one important aspect from the 
one shown in Fig. \ref{ics}. It shows a large gradient with semimajor axis with Q-type objects being rare 
beyond 1.5 AU. Conversely, the observed distribution is flat in $a$ with a significant fraction of Qs 
having $a>1.5$ AU. Some unspecified observational selection effect may be responsible for this discrepancy. 
Alternatively, this problem may indicate that the SW timescale is a function of $a$ (Marchi et al. 
2006). 

If, for example, the solar wind sputtering controls $t_{\rm sw}$ we would expect that $t_{\rm sw} 
\propto 2\pi/ \int h^{-2} = 2 q^2 (1+e)^2/(2+e^2)$, where $h$ is the heliocentric distance and 
the integral was taken over orbit. 

\begin{figure}
\epsscale{0.8}
\plotone{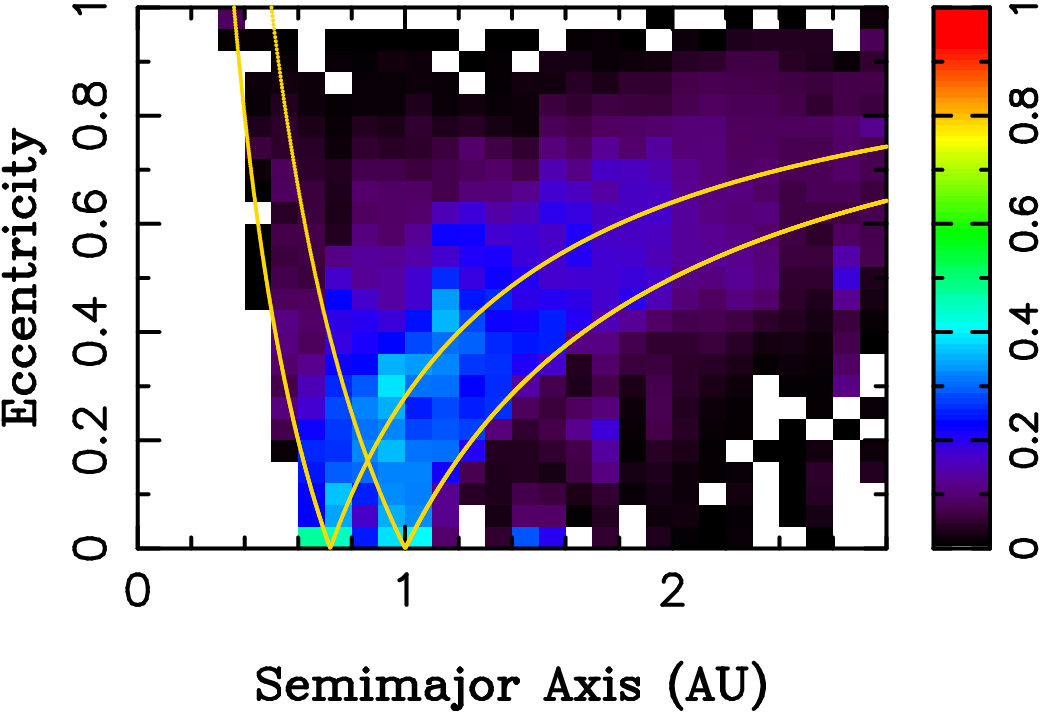}
\caption{The orbital distribution of Q-type objects expected from our NEA model with $r^*=5$ 
$R_{\rm pl}$ and $t_{\rm sw}$ that depends on heliocentric distance. Here we fixed $t_{\rm sw}=1$ My 
for a circular orbit with $a=1$ AU and increased $t_{\rm sw}$ roughly as $q^2$ with perihelion 
distance. The $(a,i)$ projection, not shown here, is very similar to Fig. \ref{m2}b.}
\label{m4}
\end{figure}

Figure \ref{m4} shows $f_Q(a,e,i)$ for $r^*=7$ $R_{\rm pl}$ and $t_{\rm sw} = 1\ {\rm My}\times 2 q^2 
(1+e)^2/(2+e^2)$ with $a$ in AU. As expected, the model distribution is flatter in $a$ with $f_Q\sim0.2$ for 
$a>1.5$ AU and $0.5<q<1$~AU. This fits observations in this orbital range rather nicely (better than our 
nominal model with fixed $t_{\rm sw}$). Note, however,
that it fails to explain 5 Q-type NEAs with $a\lesssim1$ AU and $q<0.5$ AU that represent $\sim$50\% 
of surveyed chondritic NEAs in this region (see \S5 for a discussion). The implication of 
this model is that $t_{\rm sw}>1$ My in the main asteroid belt. In \S5, we discuss how this fits the 
independent constraints obtained on $t_{\rm sw}$ from studies of asteroid families. 

\section{Discussion}

Several tidal effects may disturb the surface of a NEA during a distant planetary 
encounter.\footnote{We only discuss distant encounters here. It is clear that the SL9-like tidal disruption, 
binary formation events, or events with significant mass shedding will erase any pre-existing surface 
features.} For example:   
(1) The interior structure of a rubble-pile asteroid may find a new equilibrium by re-arranging its 
components. This motion can produce landslides, degrade craters, ballistically displace surface material, 
or even remove the original layers from the asteroid. 
(2) Tidal stresses applied to a fractured interior may produce seismic shakes similar to, or perhaps more 
effective than, those generated by impacts. Consequently, surface morphology may be modified.
(3) The tidal torque may spin up an asteroid. In surface segments where the centrifugal 
force exceeds gravity, regolith layers will be removed by carrying away the excess angular momentum. 
More subtle changes can occur in other surface parts of a spun-up asteroid.    
(4) If the tidal force becomes comparable to the object's gravity during encounter, an asteroid with 
large enough internal strength and a strengthless regolith may lose its regolith layer.    

These effects and their dependence on the encounter distance and speed are poorly understood. Some 
insights into this problem can be obtained from Richardson et al. (1998), where the authors performed 
numerical simulations of the effects of tidal gravity on a small asteroid with strengthless (rubble-pile) 
interior. In the most favorable case (slow encounter speed, fast prograde rotation), they found that 
significant mass shedding can occur up to $\approx$5~$R_{\rm pl}$. This sets a soft constraint on $r^*$. 
On one hand, $r^*$ can be larger than 5 $R_{\rm pl}$ because the optically-active thin surface 
layer may be vulnerable to even tiniest perturbations that were not considered in the Richardson et al.
model. On the other hand, when averaging over all encounter geometries and plausible asteroid spin states, 
the mean $r^*$ can become lower than 5 $R_{\rm pl}$. Thus, for the lack of additional constraints on $r^*$, 
we will tentatively assume below, as a guideline for discussion, that $r^* \sim 5$ $R_{\rm pl}$.  

If we set $r^*=5$ $R_{\rm pl}$ our results described in \S3 and \S4 imply that 
$t_{\rm sw}\sim1$~My. At first sight, this SW timescale seems to be comparable to that obtained from 
comparative studies of asteroid families in the main belt and OC meteorites in the 
RELAB database (NJWI05, Vernazza et al. 2009). For example, Vernazza et al. (2009) proposed 
that the SW timescale is $\lesssim1$ My. Their result hinges on observations of two largest members 
of the Datura family that formed by a catastrophic breakup $\approx$0.5 My ago (Nesvorn\'y et al. 2006, 
Vokrouhlick\'y et al. 2009). These two objects, 1270 Datura and 90265 2003CL5, appear to be 
significantly (but not completely) space weathered (Moth\'e-Diniz and Nesvorn\'y 2008), which implies 
that the SW timescale should be comparable to or shorter than the Datura family's age. This poses a problem 
because $t_{\rm sw} \lesssim 0.5$~My does not fit the NEA constraint (unless $r^* > 5$ $R_{\rm pl}$). 
Below we discuss possible solutions to this problem.

Observations of 2001 WY35, one of the smallest known members of the Datura family (absolute magnitude 
$H=17$), indicate that this object is not space weathered at all (Moth\'e-Diniz and Nesvorn\'y 2008). 
If these observations were correct, they would indicate that (at least some) km-sized asteroids may 
weather on timescales significantly longer than $\approx$0.5~My. For example, small km-sized fragments 
ejected from asteroid breakup events may not retain/accumulate sufficient regolith layer on their 
surface in the immediate aftermath of the collision. The SW effects may be delayed for such objects 
until a particulate (SW-sensitive) surface layer develops on their surface, for example, by subsequent 
impact shattering of the exposed rock. Thus, the regolith formation and `gardening' can be an important
part of the problem (Jedicke et al. 2004, Willman et al. 2008).

We should not forget that the two constraints on the SW timescale discussed here
come from studies of two distinct population of objects that are affected by different physical 
processes. The asteroids in the main-belt families are born by violent collisions and spend most of 
their lifetime beyond 2 AU. The NEAs, on the other hand, are exposed to more extreme solar-wind 
and temperature environment. They are olivine-rich and may therefore be more susceptible to SW
effects than an average MBA (Sasaki et al. 2001, Marchi et al. 2005).
While large impacts on NEAs should be rare, bombardment of their surface 
by $D\sim100$ $\mu$m particles should be more intense than on MBAs due to the larger number 
density of micrometeoroids at $<$2 AU (Gr\"un et al. 1985). Also, distant planetary encounters
of NEAs should produce more gentle effects than catastrophic collisions of MBAs, thus giving the 
initial surface different attributes. In summary, the {\it proper} SW timescale that measures the 
progression of SW under ideal conditions (e.g., in absence of regolith gardening) may be substantially 
shorter than the {\it apparent} SW timescale that arises from combination of different 
effects, and these effects most likely operate on different timescales in the NEA and MBA environments.

Another interesting possibility is related to the effects of the Earth magnetosphere on loose 
particulate material on a small asteroid's surface. The Earth magnetosphere extends to $\approx$12
$R_{\rm Earth}$ in the direction toward the Sun, $\approx$15 $R_{\rm Earth}$ in apex and antapex 
directions, and $\approx$25 $R_{\rm Earth}$ the anti-helion direction. The tail region stretches 
well past 200 $R_{\rm Earth}$, and the way it ends is not well-known. The magnetic field ranges 
from 30-60 $\mu$T at the Earth's surface and falls roughly as $1/r^3$ with distance $r$ toward the edge 
of the magnetosphere. Thus, if a 100-m-sized NEA passes at distance 10 $R_{\rm Earth}$, a 10-$\mu$m 
surface dust grain subject to the Lorentz force would levitate if previously charged to $>10^8$ e. Such 
charge is plausible for an asteroid surface of sufficiently high electrical resistivity. It is not clear, 
however, whether the Lorentz force effect can be more significant than the electrostatic levitation 
(Lee 1996) and/or van der Waals forces (Scheeres and Hartzell 2010).

While speculative, the effects of Earth magnetosphere could possibly allow for larger $r^*$ values than 
those expected for tidal gravity. This could perhaps help to resolve some of the discrepancy between 
different measurements of the SW timescale discussed above. For example, with $r^*=20$ $R_{\rm Earth}$, 
Fig. \ref{prob2} would imply that $t_{\rm sw} \approx$ 0.25 My. 

The orbits of Q-type NEAs in Fig. \ref{ics} hint on bimodal distribution with a group of 7 objects 
with $a\lesssim1$ AU and largely spread inclination values, and 12 objects with $a\gtrsim1.5$ AU and 
$i\lesssim10^\circ$. Using planetary encounters as the main agent that resets the SW clock, we were not
able to fit both groups simultaneously. We found that the model with fixed $t_{\rm sw}$ can match the 
low-$a$ group but it fails to fit the observed fraction of Qs with $a>1.5$ AU (e.g., Fig. \ref{m2}). 
On the other hand, the model with $t_{\rm sw} \propto q^2$ matches the high-$a$ group (Fig. \ref{m4})
but it fails to produce the observed large fraction of Qs with $a\lesssim1$ AU and $q<0.5$ AU. 

This is puzzling. The problem may be related to biases in the current sparse spectrophotometric data.
Alternatively, we may be missing some important physical effect in the model. For example, a small 
irregular object can be spun up by a radiation effect known as YORP and shed mass (e.g., Walsh et al.
2009; see Bottke et al. 2006 for a recent review of YORP). This could lead to a partial or global removal
of the space weathered material and exposure of fresh material on the surface. This effect can therefore 
be important. Unfortunately, the timescale on which the surface of a typical NEA can be reset by YORP 
is poorly understood.  

Kaasalainen et al. (2007) determined that 1862 Apollo ($a=1.47$ AU, $D=1.4$ km) is spun up by YORP on 
a characteristic timescale $t_{\rm YORP} = \omega/(d\omega/dt) \sim 2.6$ My, where $\omega$ is the spin 
rate. Starting from its current 3-hour spin period, 1862 Apollo is thus expected to be spun up to a 
$\sim$2-hour period (and start shedding mass) in $\sim$2 My. This timescale is probably at least slightly 
longer than the one on which the surface of 1862 Apollo should be reset by planetary encounters 
indicating that the YORP effect can be ignored for 1862 Apollo.

Since $t_{\rm YORP} \propto D^2 a^2 \sqrt{1-e^2}$ (e.g., Nesvorn\'y et al. 2007), however, the YORP effect 
can become more important than planetary encounters for small NEAs that orbit closer to the Sun than  
1862 Apollo. For example, a sub-km NEA with $a<1$ AU can have $t_{\rm YORP}$ several times shorter 
than 1862 Apollo. We therefore speculate that the YORP effect can contribute to the observed excess 
of Q-type NEAs in these low-$a$ orbits. A detailed analysis of this problem goes beyond the scope of 
this paper.

\section{Summary}

The main results obtained in this work can be summarized as follows:

1) The NJWI05 model (\S2) is consistent with the current spectroscopic observations of NEAs. The effect of 
planetary encounters can therefore explain the tendency towards seeing the fresh OC-like material 
among NEAs. The fraction of Q-type asteroids in the main belt should be small because the processes 
that affect MBAs (e.g., collisions) lack the efficiency of planetary encounters.

2) From modeling the spectral properties of NEAs we found that the SW timescale is
longer than $\sim$0.1 My and shorter than $\sim$10 My. It is most plausible that $t_{\rm sw}\sim1$~My 
and $r^* \sim 5$ $R_{\rm pl}$. This result is in a broad agreement with $t_{\rm sw}$ estimated 
from studies of asteroid families and our current understanding of the effects of tidal gravity.

3) We found that $t_{\rm sw} \propto q^2$, expected if the solar wind sputtering controls $t_{\rm sw}$, 
provides a better fit to the orbital distribution of Q-type NEAs than models with fixed $t_{\rm sw}$.
If $t_{\rm sw} \propto q^2$, however, our simple model fails to explain the excess of Q-type NEAs 
with low-$a$ orbits. We speculate that this population could be susceptible to the YORP effect.

4) Tidal encounters of NEAs with Venus and Earth are important, but those with Mars (and Mercury) are 
rare. This is mainly due to the fact that Mars is a much smaller planet than Venus and Earth and has a 
relatively large orbit. From the statistics of Mars encounters we estimate that a small fraction of MCAs 
could be Qs ($\lesssim1$\%). This fraction should be above the main belt average. A large observational 
sample will be needed to test this prediction.

5) The effects of the Earth's magnetosphere can be more important than tidal gravity for distant Earth 
encounters. These distant encounter effects are not required, however, to explain the observed fraction 
of Q-type NEAs, if $t_{\rm sw}\sim1$~My. 

\acknowledgements

This work was funded by the NASA Planetary Geology and Geophysics program. The work of DV was 
also partially supported by the Czech Grant Agency (grant 205/08/0064) and the Research Program 
MSM0021620860 of the Czech Ministry of Education. We thank Bruce Hapke and Robert Jedicke for
their very helpful referee reports.

\clearpage

\clearpage

\clearpage

\clearpage

\clearpage

\clearpage

\clearpage

\clearpage

\clearpage

\clearpage

\clearpage

\end{document}